\documentclass[showpacs,amsmath,amssymb,twocolumn,superscriptaddress,notitlepage,preprintnumbers,aps,pra,floatfix]{revtex4-2}

\usepackage{natbib}
\usepackage{qcircuit} 
\usepackage[dvips]{graphicx}
\usepackage{amsmath,amssymb,amsthm,mathrsfs,amsfonts,dsfont}
\usepackage{braket}
\usepackage{bm}
\usepackage{enumerate}
\usepackage{color}
\usepackage{graphicx}
\usepackage{algorithm}
\usepackage{algpseudocode}
\usepackage{braket}
\usepackage{appendix}
\usepackage{mathtools}
\usepackage{hyperref} 
\usepackage{diagbox}

\usepackage{hyperref}
\hypersetup{colorlinks=true, linkcolor=blue, citecolor=blue, urlcolor=black }

\newcommand{\tr}{\text{Tr}}

\newcommand{\mc}[1]{\mathcal{#1}}

\newcommand{\comments}[1]{}

\newcommand{\RNum}[1]{\uppercase\expandafter{\romannumeral #1\relax}}

\newcommand{\ketbra}[2]{\vert #1 \rangle \langle #2 \vert}
\newcommand{\ave}[1]{\langle #1 \rangle}

\begin{document}
\title{Experimental Hybrid Shadow Tomography and Distillation}

\author{Xu-Jie Peng}
\email{These authors contributed equally to this work.}
\affiliation{School of Physics, State Key Laboratory of Crystal Materials, Shandong University, Jinan 250100, China}

\author{Qing Liu}
\email{These authors contributed equally to this work.}
\affiliation{Key Laboratory for Information Science of Electromagnetic Waves (Ministry of Education), Fudan University, Shanghai 200433, China}

\author{Lu Liu}
\email{These authors contributed equally to this work.}
\affiliation{School of Physics, State Key Laboratory of Crystal Materials, Shandong University, Jinan 250100, China}

\author{Ting Zhang}
\affiliation{School of Physics, State Key Laboratory of Crystal Materials, Shandong University, Jinan 250100, China}

\author{You Zhou}
\email{you\_zhou@fudan.edu.cn}
\affiliation{Key Laboratory for Information Science of Electromagnetic Waves (Ministry of Education), Fudan University, Shanghai 200433, China}
\affiliation{Hefei National Laboratory, Hefei 230088, China}

\author{He Lu}
\email{luhe@sdu.edu.cn}
\affiliation{School of Physics, State Key Laboratory of Crystal Materials, Shandong University, Jinan 250100, China}
\affiliation{Shenzhen Research Institute of Shandong University, Shenzhen 518057, China}


\begin{abstract}
Characterization of quantum states is a fundamental requirement in quantum science and technology. 
As a promising framework, shadow tomography shows significant efficiency in estimating linear functions, however, for the challenging nonlinear ones, it requires measurements at an exponential cost.
Here, we implement an advanced shadow protocol, so-called hybrid shadow~(HS) tomography, to reduce the measurement cost in the estimation of nonlinear functions in an optical system. We design and realize a deterministic quantum Fredkin gate with single photon, achieving high process fidelity of $0.935\pm0.001$. Utilizing this novel Fredkin gate, we demonstrate HS in the estimations, like the higher-order moments up to 4, and reveal that the sample complexity of HS is significantly reduced compared with the original shadow protocol. Furthermore, we utilize these higher-degree functions to implement virtual distillation, which effectively extracts a high-purity quantum state from two noisy copies. The virtual distillation is also verified in a proof-of-principle demonstration of quantum metrology, further enhancing the accuracy of parameter estimation. Our results suggest that HS is efficient in state characterization and promising for quantum technologies.       
\end{abstract}
\maketitle

\noindent\textbf{\emph{Introduction.---}}Estimating properties of quantum systems is a both fundamental and practical issue for quantum science and technology \cite{Eisert2020certification,Kliesch2021Certification,gebhart2023learning}, where shadow tomography is a recently developed general framework for such important tasks~\cite{aaronson2019shadow,huang2020predicting,elben2023randomized}.
By post-processing the projective measurement results of underlying state $\rho$ on random bases, one can sequentially construct independent shadow snapshots, i.e., the unbiased estimator $\hat{\rho}$ \cite{huang2020predicting}.
Shadow tomography shows promising measurement efficiency of simultaneously estimating many linear functions, such as local observables ~\cite{huang2021efficient,hadfield2022measurements,wu2023overlapped} and quantum fidelities~\cite{elben2020cross,Zhenhuan2022correlation}, which lead to a few realizations~\cite{Struchalin2021Experimental,zhang2021experimental,Roman2022Single,Yu2021Fisher,an2023efficient}. 
On the other hand, the estimation of nonlinear functions is more challenging.
Besides many important applications from quantum information processing \cite{elben2020mixedstate,singlezhou,ketterer2019characterizing,rath2021Fisher} to many-body physics \cite{Vermersch2019Scrambling,garcia2021quantum,Elben2020topological}, nonlinear functions, such as the purity and higher-order moment $P_L=\tr(\rho^L)$, are the key ingredient in virtual distillation (VD)~\cite{Huggins2021Virtual,Koczor2021Exponential,seif2023shadow,Yamamoto2022Metrology}, playing a crucial role in error mitigation of noisy quantum processors~\cite{Brien2022purification,endo2021hybrid,cai2023quantum}. However,
the capability of the original shadow (OS) protocol for such nonlinear functions is inevitably limited by the exponentially-increasing sampling number regarding to the system size and the heavy cost of post-processing~\cite{elben2020mixedstate,singlezhou}.

A hybrid shadow (HS) tomography~\cite{zhou2022hybrid} has been proposed aiming to enhance the efficiency for estimating nonlinear functions,
which combines the advantages of shadow estimation and the generalized SWAP test~\cite{Ekert2002Direct,Horodecki2002Method}.
HS can effectively conduct randomized measurements on $\rho^k$ by introducing a coherent controlled-(generalized) SWAP operation on $k$-copy $\rho^{\otimes k}$, and therefore construct the estimator $\widehat{\rho^k}$. In this way, HS is capable of estimating higher-degree functions by patching lower-degree estimator $\widehat{\rho^k}$ via post-processing, which can significantly reduce the sampling and post-processing costs~\cite{zhou2022hybrid}. The theoretical and numerical findings there depend on the ideal multi-qubit coherent operation, and thus it remains an open question whether HS holds its effectiveness and superiority in real quantum platforms.

In this work, we give this question a positive answer by realizing HS in an optical system. We design and implement a novel quantum Fredkin gate with single photon, which is able to deterministically apply controlled-SWAP operation on two single-qubit states with high fidelity. Inspired by this novel Fredkin gate, we implement HS and show the efficiency of HS in estimation of higher-degree functions, i.e., the moment functions $P_L$ with $L=\{2,3,4\}$. We demonstrate that the measurement budget and sample complexity are significantly reduced in HS compared to OS. Moreover, we showcase the usefulness of HS in VD and error-mitigated quantum metrology. 

\begin{figure*}[t!]
    \includegraphics[width=1\textwidth]{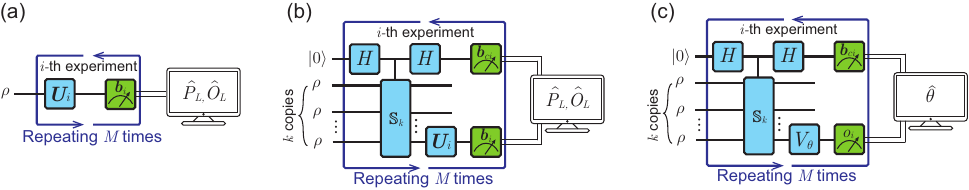}
    \caption{The scheme of (a) original shadow tomography, (b) hybrid shadow tomography, and (c)  
    error-mitigated quantum metrology
    within the hybrid framework.
    }
    \label{fig:sketch}
\end{figure*}

\noindent\textbf{\emph{Framework.---}}
We start by briefly reviewing the OS in estimating various properties of an quantum state $\rho \in \mc{H}$~\cite{huang2020predicting}. As shown in Fig.~\ref{fig:sketch}(a), one can obtain the outcome $\bm{b}$ by applying a random unitary 
operation sampled from some ensemble $\bm{U}\in\mathcal{U}$ followed with measurements on the computational basis. With the information of $\bm{U}$ and $\bm{b}$, the classical shadow is constructed as 
\begin{equation}\label{eq:os}
    \hat{\rho}=\mathcal{M}^{-1}\left(\bm{U}^\dagger\ket{\bm{b}}\bra{\bm{b}}\bm{U}\right),
\end{equation}
which is an unbiased estimator, i.e., $\mathbb{E}_{\{\bm{U},\bm{b}\}}\hat{\rho} = \rho$. Here the classical inverse $\mathcal{M}^{-1}$ is determined by the chosen random ensemble $\mathcal{U}$~\cite{huang2020predicting,hu2022Locally,ohliger2013efficient}. 

For a (linear) observable $O$, OS provides an unbiased estimation of $\ave{O}=\tr(O\rho)$ via $\hat{o}=\tr(O\hat{\rho})$. Meanwhile, OS is capable of estimating nonlinear functions, such as the higher-order moment $P_L$, and the corresponding unbiased estimator shows
\begin{equation}\label{Eq:momentOST}
\widehat{p_L}_{(\text{OS})}=\tr\left[\mathbb{S}_L\hat{\rho}_{(1)}\otimes\hat{\rho}_{(2)}\otimes\cdots\otimes\hat{\rho}_{(L)}\right]
=\tr\left[  \prod_{j=1}^L \hat{\rho}_{(j)} \right],
\end{equation}
where $\mathbb{S}_L$ denotes the generalized SWAP operation, i.e., the SHIFT operation on $\mc{H}^{\otimes L}$. 
Here, $\hat{\rho}_{(j)}$ are distinct shadow snapshots. Meanwhile, one can construct the estimator for $\tr(O\rho^L)$ as $\widehat{o_L}_{(\text{OS})}=\tr\left[ O  \prod_{j=1}^L \hat{\rho}_{(j)} \right]$.
One would enhance the estimation accuracy by repeating the experiment for $M$ times to collect the shadow set $\{\hat{\rho}_{(i)}\}_{i=1}^M$. However, the exponentially increased number of measurements with the qubit number, along with the cost for classical post-processing retards efficiency of OS in estimation of nonlinear functions~\cite{huang2020predicting,elben2020mixedstate}.

HS~\cite{zhou2022hybrid} advances OS on estimating nonlinear functions. As shown in Fig.~\ref{fig:sketch}(b), a controlled-$\mathbb{S}_k$ gate is applied on $\rho^{\otimes k}$ with the control qubit initialized as $\ket{+}$, and in general $k< L$. The measurement outcome of the control qubit in the X-basis is denoted as ${b_c}$, and one implements OS on one copy of $\rho$ with measurement outcome $\bm{b}$. For each single-shot experimental run, one can construct the estimator of $\rho^k$ as 
\begin{equation}\label{Eq:rhoa}
\widehat{\rho^k}= (-1)^{{b_c}}\cdot \mathcal{M}^{-1}\left(\bm{U}^\dagger \ket{\bm{b}}\bra{\bm{b}} \bm{U}\right),
\end{equation}
such that $\mathbb{E}_{\{U,\bm{b},b_c\}}\widehat{\rho^k} = \rho^k$, and similarly one can collect the set $\{\widehat{\rho^k}_{(i)}\}_{i=1}^{M}$. Interestingly, by ignoring the information of $b_c$ in the hybrid setting, one can also get the classical shadow of $\rho$ with $\bm{U}$ and $\bm{b}$ via Eq.~\eqref{eq:os}. (See Supplementary Materials for a detailed illustration.)

In this way, one can build estimators for many nonlinear properties. For example, $\tr(O\rho^k)$ can be directly estimated with $\widehat{o_k}=\tr(O\widehat{\rho^k})$.
Furthermore, more complex nonlinear functions with a higher degree can be estimated by combining low-degree shadow snapshots. Specifically, the unbiased estimator of $P_L$ shows
\begin{equation}\label{Eq:athmomentp}
\begin{aligned}
\widehat{p_L}_{(\text{HS})}&=\tr\left[\mathbb S_{q+r}\widehat{\rho^k}_{(1)}\otimes\cdots\widehat{\rho^k}_{(q)}\otimes\widehat{\rho}_{(1)}\otimes\cdots\otimes\hat{\rho}_{(r)}\right]\\
&=\tr\left[ \prod_{j=1}^q\widehat{\rho^k}_{(j)} \prod_{j^\prime=1}^r\hat{\rho}_{(j^\prime)}\right], 
\end{aligned}
\end{equation}
where $r=L\ (\text{mod}\ k)$ with $L=qk+r$, $\widehat{\rho^k}_{(j)}$ and $\hat{\rho}_{(j')}$ are distinct snapshots. 
Similarly, one can construct the corresponding estimator for $\tr(O\rho^L)$ as $\widehat{o_L}_{(\text{HS})}=\tr\left[ O\prod_{j=1}^q\widehat{\rho^k}_{(j)} \prod_{j^\prime=1}^r\hat{\rho}_{(j^\prime)}\right]$.
Compared to the one by OS in Eq.~\eqref{Eq:momentOST}, HS requires $q+r$ snapshots while OS requires $L$ ones, thus the classical post-processing cost for HS is significantly reduced. 
Moreover, as the estimator of HS effectively transfers an $L$-degree nonlinear function to a $(q+r)$-degree one, the statistical variance can also be reduced, thus saving the sample number of $\rho$ in practise.

\begin{figure*}[t!]
    \centering
    \includegraphics[width=1\textwidth]{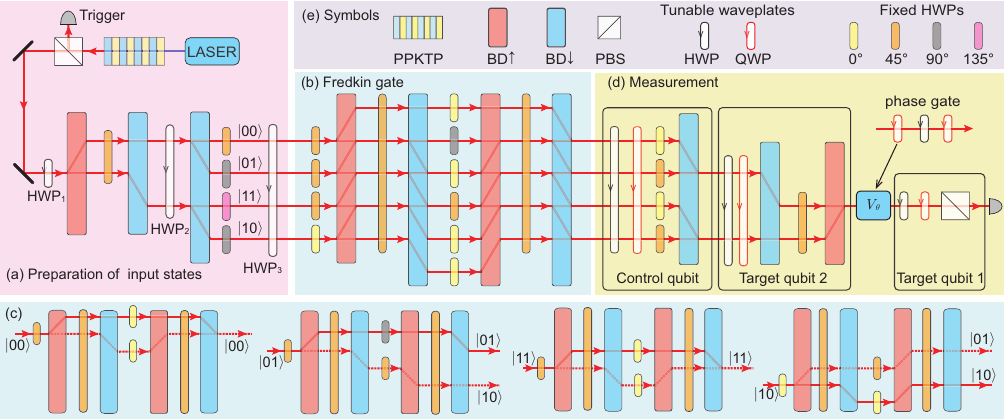}
    \caption{Schematic illustration of the experimental setup. (a) Generation of three-qubit input state $\ket{c}\ket{t_1}\ket{t_2}$, where $\ket{c}$, $\ket{t_1}$ and $\ket{t_2}$ are the states of control qubit, the first and the second target qubit respectively. (b) The optical network to implement Fredkin gate. (c) The travel path of four components $\ket{00}$, $\ket{01}$, $\ket{11}$ and $\ket{10}$ conditioned on $\ket{H}$~(dashed line) and $\ket{V}$~(solid line) respectively. (d) The setup to implement projective measurement on three qubits. (e) Symbols used in (a), (b), (c) and (d). PPKTP: periodically poled potassium titanyl phosphate; BD$\uparrow$: beam displacer that transmits $\ket{V}$ and upward deviates $\ket{H}$; BD$\downarrow$: beam displacer that transmits $\ket{V}$ and downward deviates $\ket{H}$; PBS: polarizing beam splitter that transmits $\ket{H}$ and reflects $\ket{V}$; HWP: half-wave plate; QWP: quarter-wave plate. }
    \label{fig:setup}
\end{figure*}

Apart from many applications~\cite{neven2021symmetry,Yu2021Optimal,liu2022detecting}, nonlinear functions are vital in purification-based quantum error mitigation, say VD~\cite{Huggins2021Virtual,Koczor2021Exponential,Yamamoto2022Metrology}. VD enables to approach the noiseless value of the observable $O$ over noisy state $\rho$ by $\langle O \rangle^{(L)}_\text{VD} = \tr\left(O\rho^L\right) / \tr\left( \rho^L \right)$, which is more accurate than the noisy value $\tr( O\rho)$. Such enhancement increases exponentially with the degree $L$. VD benefits a few quantum information tasks, such as quantum metrology \cite{Yamamoto2022Metrology,kwon2024efficacy}.
In quantum metrology \cite{giovannetti2004quantum,giovannetti2011advances}, a probe state $\ket{\psi}$ evolves under the dynamics $V_\theta$, and the estimator $\hat{\theta}$ of the unknown parameter $\theta$ can be constructed via measuring the evolved state $\ket{\psi_\theta}$ using some observable $O$. A noisy preparation $\rho$ instead of $\ket{\psi}$ would introduce considerable systematic errors \cite{Yamamoto2022Metrology,takeuchi2019quantum,okane2021quantum}. Here in the hybrid framework, we follow the spirit of VD to feed one of the output port of $\rho^{\otimes k}$ \emph{after} the control gate to the evolution $V_\theta$ and finally make the measurement in the eigenbasis of some observable $O$, as shown in Fig.~\ref{fig:sketch}(c). After repeating the experiment, $\langle O \rangle^{(k)}_\text{VD}$ can be estimated by
\begin{equation}\label{eq:o2_ave}
    \frac{\sum_{i\in[M]}(-1)^{{b_c}_i}o_i/M}{\sum_{i'\in[M']}(-1)^{{b_c}_{i'}}/{M^\prime}},
\end{equation} 
where $o_i$ is the outcome of $O$ for the $i$-th shot. Such approach significantly enhances the estimating accuracy, as it effectively realizes quantum metrology with the purified state $\rho^k/\tr(\rho^k)$, which is much closer to the ideal one $\ket{\psi}$ than $\rho$. We should remark that our scheme only needs to feed the purified port to the environment (other than all $k$ copies) and do \emph{not} require the controlled-$O$ operation compared to Ref.~\cite{Yamamoto2022Metrology}.
(See Supplementary Materials for proofs and more details.)

\noindent\textbf{\emph{Experiment.---}} We implement the HS using a novel quantum Fredkin gate. In contrast to multi-photon Fredkin gate~\cite{Raj2016SA,Ono2017SR,Li2022NPJQI}, our Fredkin gate acts on three qubits encoded on different degrees of freedom~(DOF) of a single photon. The single-photon Fredkin gate is deterministic and generally can achieve high fidelity as it does not depend on multi-photon interference and post-selection~\cite{Wang2021QST}. In our Fredkin gate, the control qubit is encoded in the polarization DOF $\ket{H}=\ket{1}$ and $\ket{V}=\ket{0}$, with $\ket{H}$~($\ket{V}$) being the horizontal~(vertical) polarization. Two target qubits are equivalently encoded in a four-dimensional Hilbert space spanned by the four path modes $\ket{l_0}=\ket{00}$, $\ket{l_1}=\ket{01}$, $\ket{l_2}=\ket{11}$ and $\ket{l_3}=\ket{10}$. As illustrated in Fig.~\ref{fig:setup}(a), the heralded single photon is obtained by triggering one photon of photon pair generated from a periodically poled potassium titanyl phosphate crystal via spontaneous parametric down conversion. We use three tunable half-wave plates~(HWP$_1$, HWP$_2$ and HWP$_3$), four HPWs with fixed angles and three beam displacers~(BDs) to prepare the input state $\ket{c}\ket{t_1}\ket{t_2}$~(hereafter referred as $\ket{ct_1t_2}$), where $\ket{c}=\cos2\theta_3\ket{H}+\sin2\theta_3\ket{V}$, $\ket{t_2}=\cos2\theta_2\ket{0}+\sin2\theta_2\ket{1}$ and $\ket{t_1}=\cos2\theta_1\ket{0}+\sin2\theta_1\ket{1}$. $\theta_1$, $\theta_2$ and $\theta_3$ are the angles of HWP$_1$, HWP$_2$ and HWP$_3$ respectively. See Supplementary Materials for details of state preparation. The Fredkin gate is realized with the optical network shown in Fig.~\ref{fig:setup}(b). To give a clear illustration, we individually show how the four path modes $\ket{00}$,$\ket{01}$,$\ket{10}$ and $\ket{11}$ pass through the optical network conditioned on $\ket{H}$~(solid line) and $\ket{V}$~(dashed line) in Fig.~\ref{fig:setup}(d). The projective measurement is sequentially performed on the control qubit, the second and the first target qubit as shown in~Fig.~\ref{fig:setup}(c). (See Supplementary Materials for details of projective measurement.) 

\begin{figure*}[ht!p]
    \centering    \includegraphics[width=\linewidth]{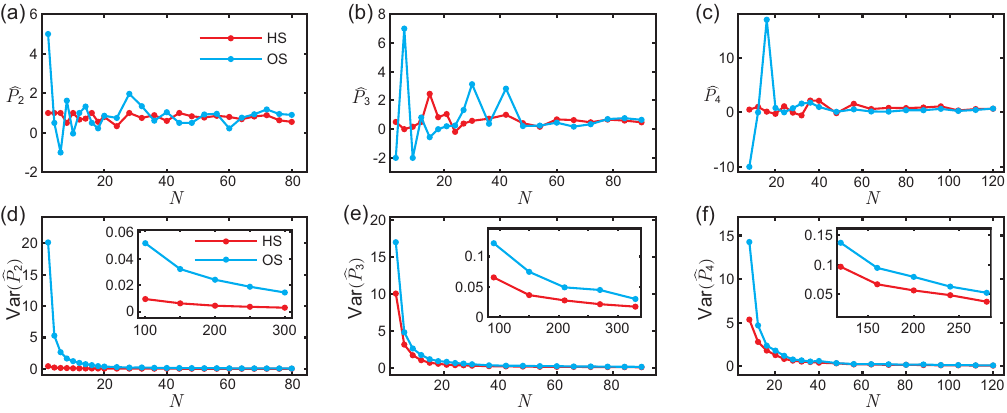}
    \caption{Experimental results of estimations (a) $\widehat{P_{2}}$, (b) $\widehat{P_{3}}$ and (c) $\widehat{P_{4}}$. (d)-(f) The variances of estimations $\widehat{P_{2}}$, $\widehat{P_{3}}$ and $\widehat{P_{4}}$. The red and blue dots represent the results with HS and OS respectively, and the x-axis is the consumed number of the state $\rho$, labeled by $N$.
    }
    \label{fig:nonlinear}
\end{figure*}

We first characterize the Fredkin gate by measuring the truth table, according to which we observe the classical fidelity of $0.9898\pm0.0003$ with respect to the ideal case. The Fredkin gate is also capable to generate three-qubit Greenberger-Horne-Zeilinger~(GHZ) state $\ket{\text{GHZ}}=(\ket{010}+\ket{101})/\sqrt{2}$ by inputting $\ket{+10}$. The fidelity of generated GHZ state is $0.933\pm0.002$. Furthermore, we estimate the process fidelity $0.935\pm0.001$ of the implemented gate. (See Supplementary Materials for more experimental results regarding the characterization of Fredkin gate.)

We implement HS on two copies of single-qubit noisy state $\rho=0.8\ket{+}\bra{+}+0.2\mathbb{I}/2$, where the two-qubit target state $\rho^{\otimes 2}$ can be written as 
\begin{equation}\label{Eq:twocopy}
\begin{aligned}
\rho^{\otimes 2}=&0.64\ket{++}\bra{++}+0.08\ket{+1}\bra{+1}+0.08\ket{+0}\bra{+0}\\
&0.08\ket{1+}\bra{1+}+0.01\ket{11}\bra{11}+0.01\ket{10}\bra{10}\\
&0.08\ket{0+}\bra{0+}+0.01\ket{01}\bra{01}+0.01\ket{00}\bra{10},
\end{aligned}   
\end{equation}
and can be prepared by randomly setting angles $[\theta_1, \theta_2, \theta_3]$ to generate $\ket{t_1t_2}\in\{\ket{++}$, $\ket{+1}$, $\ket{+0}$, $\ket{1+}$, $\ket{11}$, $\ket{10}$, $\ket{0+}$, $\ket{01}$,  $\ket{00}\}$, with probability of 0.64, 0.08, 0.08, 0.08, 0.01, 0.01, 0.08, 0.01, 0.01~(See supplementary Materials for the angle settings.) After the Fredkin gate, the control qubit is measured on Pauli-$X$ basis. The target qubit is randomly detected on Pauli-$X$, $Y$ and $Z$ bases. By running the experiment many times, the unbiased estimators of $P_2$, $P_3$ and $P_4$ are constructed by 
\begin{equation}\label{eq:HSP234}
    \begin{split}
    &\widehat{P_2}_{(\text{HS})}=\frac{1}{M}\sum_{i\in \left[ M \right]}{\tr\left[ \widehat{\rho^2} _{(i)} \right]},\\ 
    &\widehat{P_3}_{(\text{HS})}= \frac{1}{MM^\prime}\sum_{i\in[M],i^\prime\in[M^\prime]}{\tr\left[\widehat{\rho^2}_{(i)}\hat{\rho}_{(i^\prime)} \right]},\\
    &\widehat{P_4}_{(\text{HS})}= \frac{1}{MM^\prime}\sum_{i\in[M],i^\prime\in[M^\prime]}{\tr\left[ \widehat{\rho^2} _{(i)}\widehat{\rho^2}_{(i^\prime)} \right]},
    \end{split}
\end{equation}
according to Eq.~\eqref{Eq:athmomentp},
with $\{\hat{\rho}_{(i)}\}$ and $\{\widehat{\rho^2}_{(i)}\}$ constructed via Eq.~\eqref{eq:os} and \eqref{Eq:rhoa}, respectively. The experimental results of $\widehat{P_{2}}$, $\widehat{P_{3}}$ and $\widehat{P_{4}}$ are shown with red dots in Fig.~\ref{fig:nonlinear}(a)-Fig.~\ref{fig:nonlinear}(c), respectively. To give a comparison, we also implement OS to estimate $\widehat{P_{2}}$, $\widehat{P_{3}}$ and $\widehat{P_{4}}$ according to Eq.~\eqref{Eq:momentOST}, and the results are shown with blue dots in Fig.~\ref{fig:nonlinear}(a)-Fig.~\ref{fig:nonlinear}(c) respectively. Note that the HS consumes two copies of $\rho$ in each experiment run while the OS consumes one, so that the comparison is carried out with the same number of consumed copies (denoted by $N$) instead of experimental runs $M$. As reflected in Fig.~\ref{fig:nonlinear}(a)-Fig.~\ref{fig:nonlinear}(c), the results of HS converge faster than that of OS, which clearly demonstrate the reduction of sample complexity using the hybrid framework. 
\begin{figure*}[t!]
    \centering
    \includegraphics[width=1\textwidth]{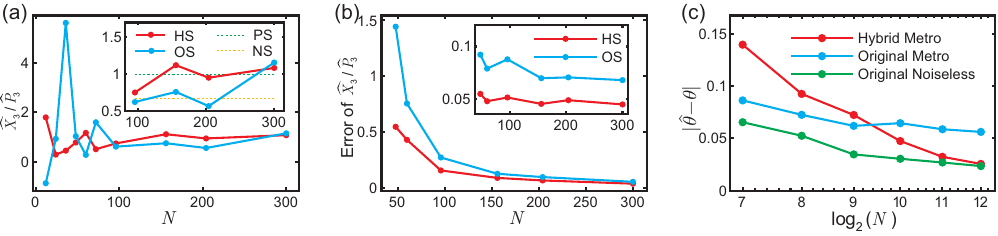}
    \caption{(a) Experimental results of VD of the expectation value of observable Pauli $X$, with the estimator 
$\widehat{X_3}/\widehat{P_3}$, for both HS and OS. The green and yellow dashed lines in insert represent the theoretical prediction of $\langle X\rangle$ on $\ket{+}$ and noisy $\rho$ respectively. (b) Estimation error 
of $\widehat{X_3}/\widehat{P_3}$. (c) The accuracy of the estimation $\hat{\theta}$ using HS~(red dots) and original single-copy method~(blue dots); The green dots represent the results with (almost) perfect pure state $\ket{+}$.}
    \label{fig:moments}
\end{figure*}

The sample complexity is further investigated by the variance of $\widehat{P_{L}}$~\cite{huang2020predicting,innocenti2023shadow}
\begin{equation}
    \text{Var}\left( \widehat{P_{L}}\right) = \mathbb{E}\left( \widehat{P_{L}}-P_L\right)^2. 
\end{equation}
Note that Var$(\widehat{P_{L}})\to0$ when $N\to\infty$. In the experiment, the converged value $P_L$ is estimated by using a large number~($3\times10^4$) of copies. The results of Var$(\widehat{P}_2)$, Var$(\widehat{P}_3)$, Var$(\widehat{P}_4)$ are shown in Fig.~\ref{fig:nonlinear}(d)-Fig.~\ref{fig:nonlinear}(f) respectively, and we observe Var$(\widehat{P_{L}})$ of HS converges to zero faster than that of OS.

Next, we show VD of the expectation value of some observable $O$ with the assistance of $\widehat{P_{L}}$ and $\widehat{O_L}$. That is, estimating $\langle O \rangle^{(L)}_\text{VD}$ with $\widehat{O_L} /\widehat{P_{L}}$, where $\widehat{O_L}$ is an estimator of $\tr(O\rho^L)$. For example, $\widehat{O_3}$ for HS is constructed as
\begin{equation}
    \widehat{O_3}_{(\text{HS})}=\frac{1}{MM^\prime}\sum_{i\in \left[ M \right] ,i^{\prime}\in \left[ M^{\prime} \right]}{\tr\left[ O \widehat{\rho ^2}_{ (i) }\widehat{\rho }_{ (i^{\prime}) }  \right]},
\end{equation}
in a similar way as $\widehat{P_3}_{(\text{HS})}$ in Eq.~\eqref{eq:HSP234}.
The result of estimating $\langle X \rangle^{(3)}_\text{VD}$ for observable $O=X$ is shown in Fig.~\ref{fig:moments}(a). After VD, $\langle X \rangle^{(3)}_\text{VD}$ is closer to 1, which is the theoretical prediction of $\langle X\rangle$ on state $\ket{+}$~( green dashed line in insert of Fig.~\ref{fig:moments}(a)). Similarly, the variance of $\langle X \rangle^{3}_\text{VD}$ converges faster compared to OS as reflected in Fig.~\ref{fig:moments}(b). See Supplementary Materials for the results of $\langle X \rangle^{(2)}_\text{VD}$ and $\langle X \rangle^{(4)}_\text{VD}$.  Note that $\langle X \rangle^{(4)}_\text{VD}$ is more accurate than $\langle X \rangle^{(3)}_\text{VD}$.

Finally, we demonstrate quantum metrology with the assistance of VD in such a hybrid framework. We consider the probe qubit passing through a phase gate ${V_\theta}=e^{-i\frac{\theta}{2}Z}$ as shown in Fig.~\ref{fig:setup}(e), and the value of $\theta$ is the parameter to be estimated. By measuring the probe qubit in the eigenbasis of Pauli $Y$, one can obtain the estimation $\hat{\theta}=\widehat{Y_L}/\widehat{P_L}$. The optimal accuracy is achieved with the perfect probe state of $\ket{+}$. In our experiment, we set $\theta=\pi/15$ and the initial probe state is a noisy state  $\rho=0.8\ket{+}\bra{+}+0.2\mathbb{I}/2$. Without VD, $\delta\hat{\theta}$ is illustrated with blue dots in Fig.~\ref{fig:moments}(c). The results of $\delta\hat{\theta}$ with VD are shown with red dots in Fig.~\ref{fig:moments}(c), and we observe $\delta\hat{\theta}$ with VD is smaller than that without it when around $2^{10}$ copies of $\rho$ are consumed in the experiment. Also, we perform the metrology with perfect probe state of $\ket{+}$, and the results are shown with green dots in Fig.~\ref{fig:moments}(c). The accuracy of $\hat{\theta}$ with noisy state $\rho$ achieves that with (almost) pure state $\ket{+}$ when the virtual distillation is performed with around $2^{12}$ copies.         

\noindent\textbf{\emph{Conclusion.---}}
In this work, we design and implement a deterministic and high-fidelity quantum Fredkin gate with single photon, and demonstrate the advantages of HS in important tasks like quantum error mitigation and quantum metrology. Our protocols in principle can be systemically generalized to larger systems by using parallel Fredkin gates and entangled control qubits. Note that the Fredkin gate here would benefit the investigation of other photonic quantum information tasks, such as quantum fingerprinting~\cite{Buhrman2001PRL} and optimal cloning~\cite{Hofmann2012PRL}. We also expect HS can be further applied to other important scenarios in quantum information science, like quantum algorithm design \cite{Lubasch2020nonlinear,McArdle2020chemistry}, quantum chaos diagnosis \cite{Yoshida_2019,Leone2021quantumchaosis}, and quantum magic-resource quantification \cite{leone2022stabilizer,chen2023magic,oliviero2022measuring}.
The extension of HS to the multi-shot scenario \cite{helsen2023thrifty,zhou2023performance} and continuous-variable quantum systems \cite{becker2024classical,gandhari2024precision} are also promising. Our research significantly enriches the scope of randomized measurements and shadow tomography, and advances its real-world experimental implementations, especially in the estimation of highly nonlinear functions.

\bibliographystyle{apsrev4-2}
\bibliography{main_v6.bbl}
\newpage
\appendix
\onecolumngrid
\section{Additional notes for the hybrid framework}
Here, we prove that one can construct the unbiased estimator of $\rho$ with the information $\bm{U}$ and $\bm{b}$ from the hybrid setting as shown in Fig.~1(b) in main text via
\begin{equation}
    \hat{\rho}=\mathcal{M}^{-1}(\bm{U^\dag\ket{\bm{b}}\bra{\bm{b}}U}),
\end{equation}
by ignoring the information of $b_c$. 
\begin{proof}
The expectation value of the estimator on random unitary $\bm{U}$ and the measurement result $b_c$, $\bm{b}$ is
\begin{equation}\label{eq:exp_rhohat}
\begin{aligned}
    \mathbb{E}_{\{\bm{U},b_c,\bm{b}\}}\hat{\rho} &=\mathbb{E}_{\{\bm{U},b_c,\bm{b}\}}\mathcal{M}^{-1}(\bm{U^\dag\ket{\bm{b}}\bra{\bm{b}}U})\\
    &=\sum_{\bm{U},b_c,\bm{b}}\Pr(\bm{U},b_c,\bm{b})\mathcal{M}^{-1}(\bm{U^\dag\ket{\bm{b}}\bra{\bm{b}}U})\\
    &=\sum_{\bm{U}}\Pr(\bm{U})\sum_{b_c,\bm{b}}\Pr(b_c,\bm{b}|\bm{U})\mathcal{M}^{-1}(\bm{U^\dag\ket{\bm{b}}\bra{\bm{b}}U}).
\end{aligned}
\end{equation}
Given $\bm{U}$ and $\bm{b}$, one can sum the index $b_c$ as
\begin{equation}
\begin{aligned}
    \sum_{b_c}\Pr(b_c,\bm{b}|\bm{U})=&\tr\left[\bra{+}_c\bra{\bm{b}}\bm{U}\ CS_k\left(\ket{+}_c\bra{+}_c \otimes \rho^{\otimes k}\right)CS_k^\dag \bm{U}^\dag\ket{\bm{b}}\ket{+}_c \right]\\
    &+\tr\left[\bra{-}_c\bra{\bm{b}}\bm{U}\ CS_k\left(\ket{+}_c\bra{+}_c \otimes \rho^{\otimes k}\right)CS_k^\dag \bm{U}^\dag\ket{\bm{b}}\ket{-}_c \right]\\
    =&\tr\left[ \mathbb{I}_c \bra{\bm{b}}\bm{U}\ CS_k\left(\ket{+}_c\bra{+}_c \otimes \rho^{\otimes k}\right)CS_k^\dag \bm{U}^\dag\ket{\bm{b}}\right]\\
    =&\tr(\rho)^{k-1}\bra{\bm{b}}\bm{U}\rho \bm{U}^\dag\ket{\bm{b}} = \bra{\bm{b}}\bm{U}\rho \bm{U}^\dag\ket{\bm{b}},
\end{aligned}
\end{equation}
with $CS_k$ denoting the controlled-$\mathbb{S}_k$ gate.
Thus, one can rewrite Eq.~\eqref{eq:exp_rhohat} as
\begin{equation}
    \mathbb{E}_{\{\bm{U},b_c,\bm{b}\}}\hat{\rho} = \sum_{\bm{U}}\sum_{\bm{b}}\Pr(\bm{U})\bra{\bm{b}}\bm{U}\rho \bm{U}^\dag\ket{\bm{b}}\mathcal{M}^{-1}(\bm{U^\dag\ket{\bm{b}}\bra{\bm{b}}U})=\rho.
\end{equation}
This equation is based on the definition of unbiased estimator given in the original shadow estimation protocol~\cite{huang2020predicting}.
\end{proof}

As introduced in main text, we implement OS tomography on one copy of $\rho$ after the SWAP-test. In fact, one can perform randomized measurements on each copy of $\rho$ in $\rho^{\otimes k}$ and record the corresponding $\{\bm{U}, \bm{b}\}$. 
Therefore, one can build $k$ distinct shadow snapshots for $\hat{\rho}$ with the information of $\{\bm{U}, \bm{b}\}$ for different copies in single-shot. This strategy would save the sample number of $\rho$ when constructing the shadow set $\{\hat{\rho}_{(i)}\}$, thus reduce the sample complexity. 
We applied such scheme in the classical post-processing procedure in our experiments.

\section{Quantum metrology with the hybrid framework}
\subsection{Variance of estimating $\theta$ with noisy state}

In general quantum metrology settings \cite{giovannetti2004quantum,giovannetti2011advances}, one prepares a probe state $\rho=\ketbra{\psi}{\psi}$ and evolves the state under the dynamics $V_\theta$ characterized by parameter $\theta$. Then, one performs projective measurements described by $O$ on this state and records the outcomes $o_i\in\{1,-1\}$. By fitting the experimental outcomes with a theoretical estimator $\ave{o}=\tr(O\rho_\theta)$ with $\rho_\theta = V_\theta \rho V_\theta^\dag$, one can estimate the parameter $\theta$.

Here, we discuss in detail the variance of estimating $\theta$ with the theoretical $\langle o\rangle = \tr(O\rho_\theta)=a+b\theta$~\cite{Yamamoto2022Metrology}. First, we derive the theoretical $a$ and $b$ for the ideal case when we set $V_\theta = e^{-i\frac{\theta}{2}Z}$ and measure the evolved state in y-basis, i.e., $O=Y$, as in the experiment. Here, we denote the ideal initial probe state as $\rho$, thus
\begin{equation}
\begin{aligned}
    \langle y \rangle &= \tr(Y\rho_\theta) = \tr\left[  Y\left(\cos\frac{\theta}{2}I-i\sin\frac{\theta}{2}Z\right)\rho\left(\cos\frac{\theta}{2}I+i\sin\frac{\theta}{2}Z \right) \right]\\
    &=\frac{1+\cos\theta}{2}\tr(Y\rho) + i\frac{\sin\theta}{2}\tr(Y\left[\rho,Z\right])+\frac{1-\cos\theta}{2}\tr(YZ\rho Z)\\
    &=\cos\theta\tr(Y\rho)+\sin\theta\tr(X\rho).
\end{aligned}
\end{equation}
When $\theta$ is small enough, one can ignore the higher order terms of $\theta$, that is 
\begin{equation}
    \langle y\rangle_{\theta\to 0} =\tr(Y\rho)+\tr(X\rho)\theta.
\end{equation}
Then, one can use the theoretical $a=\tr(Y\rho)$, $b=\tr(X\rho)$ to further estimate $\theta$, where the ideal (or theoretical) estimator of $\theta$ is constructed as $\hat{\theta}=(\hat{y_\theta}-a)/b$, with $\hat{y_\theta}$ the variable of the average measurement outcome $y_i\in\{1,-1\}$ characterizing $\langle y^\prime\rangle=\tr(Y\rho^\prime_\theta)$. 

In practise, the real initial quantum state may be a noisy one denoted as $\rho^\prime$. Thus, the real estimator of $\theta$ should be defined as $\hat{\theta^\prime}=(\hat{y_\theta}-a^\prime)/b^\prime$, with $a^\prime = \tr(Y\rho^\prime)$ and $b^\prime=\tr(X\rho^\prime)$. However, due to a lack of information about the noisy state $\rho^\prime$, one cannot access the real $a^\prime$ and $b^\prime$. Thus, the variance of estimating $\theta$ with the theoretical estimator follows
\begin{equation}\label{eq:app_var}
\begin{aligned}
    \text{Var}(\hat{\theta})&=\ave{(\hat{\theta}-\ave{\hat{\theta^\prime}})^2} = \ave{\hat{\theta}^2}-\ave{2\hat{\theta}\ave{\hat{\theta^\prime}}}+\ave{\hat{\theta^\prime}}^2\\
    &=\frac{1}{b^2}\mathbb{E}\left(\hat{y_\theta}^2-2a\hat{y_\theta}+a^2\right) + \frac{2}{bb^\prime}\left[(\mathbb{E}\hat{y_\theta})^2-(a+a^\prime)\mathbb{E}\hat{y_\theta}+aa^\prime\right] + \frac{1}{{b^\prime}^2}\mathbb{E}\left(\hat{y_\theta}^2-2a^\prime\hat{y_\theta}+{a^\prime}^2\right)\\
    &=\left( \frac{1}{b^2}\mathbb{E}(\hat{y_\theta}^2)-\left(\frac{2}{bb^\prime}-\frac{1}{{b^\prime}^2}\right)\left(\mathbb{E}\hat{y_\theta}\right)^2 \right) + \left( -\frac{2a}{b^2}+\frac{2(a+a^\prime)}{bb^\prime}-\frac{2a^\prime}{{b^\prime}^2} \right)\mathbb{E}\hat{y_\theta}  +  \left( \frac{a^2}{b^2}-\frac{2aa^\prime}{bb^\prime}+\frac{{a^\prime}^2}{{b^\prime}^2} \right).
\end{aligned}
\end{equation}
Note that for the ideal case when $\rho^\prime = \rho$, the variance of estimating $\theta$ follows
\begin{equation}
    \text{Var}(\hat{\theta}) = \frac{1}{b^2}\left[\mathbb{E}(\hat{y_\theta}^2)-(\mathbb{E}\hat{y_\theta})^2\right]=\frac{1}{b^2}\text{Var}(\hat{y_\theta}).
\end{equation}
Such estimation variance arises from the statistical error of the variable $\hat{y_\theta}$, which decreases with the increasing of $M$. 

Therefore, when $\rho^\prime\neq \rho$, as illustrated in Eq.~\eqref{eq:app_var}, the variance of estimating $\theta$ with the original metrology setting suffers from both statistical and systematic error \cite{Yamamoto2022Metrology,takeuchi2019quantum,okane2021quantum}.
Specifically, the first term of Eq.~\eqref{eq:app_var} is mainly generated by the statistical error; while other terms describe the systematic error due to the incorrect estimation of $\ave{y^\prime}$, that is, the value of $a^\prime$ and $b^\prime$. Note that, the first term is also influenced by the systematic error when $b^\prime \neq b$.

In main text, we consider the error of estimating $\theta$ described by $\delta\hat{\theta}=|\hat{\theta}-\theta|$ for simplicity.
Note that the error contains both statistical and systematic error. Thus, one should adopt error mitigation technique to reduce the systematic the error, which leads to our hybrid framework.

\subsection{Error-mitigated quantum metrology} 
In this part, we introduce the hybrid framework for quantum metrology described in Fig. 1(c) and prove the unbiasedness of the estimator described in Eq.(5) in main text.

The hybrid framework for quantum metrology follows the spirit of VD. Specifically, one prepares $k$ copies of the probe state $\rho$ and sends them to the generalized SWAP test circuit, with the measurement outcome of the control qubit as ${b_c}_{i}$. Next, one feeds one of the output port of $\rho^{\otimes k}$ after the controlled-$\mathbb{S}_k$ to the evolution $V_\theta$ and makes measurement in the eigenbasis of observable $O$. The outcome is denoted as $o_i\in\{1,-1\}$. After repeating the experiment, one can estimate $\ave{O}^{(k)}_{\text{VD}}$ with ${b_c}_i$ and $o_i$ via Eq.(5) in main text, that is
\begin{equation}\label{eq:supp_metro}
    \frac{\sum_{i\in[M]}(-1)^{{b_c}_i}o_i/M}{\sum_{i'\in[M']}(-1)^{{b_c}_{i'}}/{M^\prime}},
\end{equation} 
which is an unbiased estimator of $\ave{O}^{(k)}_{\text{VD}}$. With this value, one can further estimate parameter $\theta$ in $V_\theta$ with higher precision using the metrology technique. 

\begin{proof} 
Here, we prove that $(-1)^{{b_c}_{i}}o_i$ in the numerator term of Eq.~\eqref{eq:supp_metro} is an unbiased estimator of $\tr(O\rho_\theta^k)$. 
The Hermitian operator $O$ can be decomposed as $O = \sum o_{\bm{b}}\ket{\bm{b}}\bra{\bm{b}}$, with $o_{\bm{b}}$ the eigenvalue, which also describes the measurement outcome while measuring in the eigenbasis $\ket{\bm{b}}$ of $O$.
The joint density matrix for the whole system before all the projective measurements is 
\begin{equation}
\begin{aligned}
    \rho_{c,k}&=\frac{1}{2}\ketbra{0}{0}\otimes(\mathbb{I}^{\otimes k-1}\otimes V_\theta)\rho^{\otimes k} (\mathbb{I}^{\otimes k-1}\otimes V_\theta^\dag)+\frac{1}{2}  \ketbra{0}{1}\otimes(\mathbb{I}^{\otimes k-1}\otimes V_\theta)\rho^{\otimes k}\mathbb{S}_k (\mathbb{I}^{\otimes k-1}\otimes V_\theta^\dag)\\
    &+\frac{1}{2}\ketbra{1}{0}\otimes(\mathbb{I}^{\otimes k-1}\otimes V_\theta)\mathbb{S}_k\rho^{\otimes k} (\mathbb{I}^{\otimes k-1}\otimes V_\theta^\dag)+\frac{1}{2}\ketbra{1}{1}\otimes(\mathbb{I}^{\otimes k-1}\otimes V_\theta)\mathbb{S}_k\rho^{\otimes k}\mathbb{S}_k (\mathbb{I}^{\otimes k-1}\otimes V_\theta^\dag).
\end{aligned}
\end{equation}
When the measurement outcomes for one of the $\rho^{\otimes k}$ is $o_{\bm{b}}$, the density matrix of the control qubit is
\begin{equation}
    \rho_c = \frac{1}{2}\bra{\bm{b}} V_\theta\rho V_\theta^\dag\ket{\bm{b}}\mathbb{I}_c + \frac{1}{2}\bra{\bm{b}} V_\theta\rho^k V_\theta^\dag\ket{\bm{b}}X_c.
\end{equation}
Thus, when measuring the control qubit in $x$-basis, the probabilities for outcome $0$ and $1$ are
\begin{equation}
\begin{aligned}
    \Pr(0,\bm{b}) &= \bra{+}\rho_c\ket{+} = \frac{1}{2}\bra{\bm{b}}V_\theta\rho V_\theta^\dag\ket{\bm{b}} + \frac{1}{2}\bra{\bm{b}}V_\theta\rho^k V_\theta^\dag\ket{\bm{b}} \\
    \Pr(1,\bm{b}) &= \bra{-}\rho_c\ket{-} = \frac{1}{2}\bra{\bm{b}}V_\theta\rho V_\theta^\dag\ket{\bm{b}} - \frac{1}{2}\bra{\bm{b}}V_\theta\rho^k V_\theta^\dag\ket{\bm{b}},
\end{aligned}
\end{equation}
respectively. Thus, in the $i$-th shot, the expectation value shows (we ignore the subscript $i$ in ${b_c}_i$ and $\bm{b}_i$ for simplicity)
\begin{equation}\label{eq:unbiased_o}
\begin{aligned}
    \mathbb{E}_{\{{b_c},\bm{b}\}}(-1)^{b_c}o_{\bm{b}} &= \sum_{b_c,\bm{b}}\Pr(b_c,\bm{b})(-1)^{b_c}o_{\bm{b}}\\
    &=\sum_{\bm{b}}o_{\bm{b}} \left[\Pr(0, \bm{b}) - \Pr(1,\bm{b}) \right]\\
    &=\sum_{\bm{b}}o_{\bm{b}}\bra{\bm{b}}V_\theta \rho^k V_\theta^\dag \ket{\bm{b}}\\
    & =\sum_{\bm{b}}o_{\bm{b}}\bra{\bm{b}}\rho_\theta^k \ket{\bm{b}} = \tr(O\rho_\theta^k).
\end{aligned}
\end{equation}
Here, we use the property that $\rho_\theta^k = (V_\theta\rho V_\theta^\dag)^k = V_\theta\rho^k V_\theta^\dag$ in the last line of Eq.~\eqref{eq:unbiased_o}.
\end{proof}

Note that, when $O=\mathbb{I}$, we can prove that $\mathbb{E}_{\{b_c,\bm{b}\}}(-1)^{b_c} = \tr(\rho_\theta^k)$, which shows that $(-1)^{{b_c}_i}$ in the denominator term of Eq.~\eqref{eq:supp_metro} is an unbiased estimator of $\tr(\rho_\theta^k)$.

\section{Details of experiment }

We start by presenting an overview of the critical optical components utilized in our experimental setup and elucidate their respective functions:
\begin{enumerate}
    \item Waveplates: Central to our experimental design are HWPs and QWPs. The angle parameters, $\theta$ for HWPs and $\zeta$ for QWPs, define the orientation of the fast axis relative to the vertical polarization direction.
    
- HWP Unitary Transformation ($U_{\text{HWP}}(\theta)$):
    \[
    U_{\text{HWP}}(\theta) =- \begin{pmatrix}
    \cos 2\theta & \sin 2\theta \\
    \sin 2\theta & -\cos 2\theta \\
    \end{pmatrix}
    \]

    - QWP Unitary Transformation ($U_{\text{QWP}}(\zeta)$):
    \[
    U_{\text{QWP}}(\zeta) = \frac{1}{\sqrt{2}}\begin{pmatrix}
    1+i\cos 2\zeta & i\sin 2\zeta \\
    i\sin 2\zeta & 1-i\cos 2\zeta \\
    \end{pmatrix}
    \]
    \item Beam displacer~(BD): A BD transmits the vertically polarized photons and deviates the horizontally polarized photons.
    \item Polarizing Beam Splitter~(PBS): A PBS transmits the horizontal polarization and reflects the vertical polarization.
\end{enumerate}

\subsection{Preparation of input states}

We use an ultraviolet diode laser to pump a periodically poled potassium titanyl phosphate
crystal. The generated two photons are with orthogonal polarizations, and are separated by a polarizing beam splitter. The reflected photon is detected by a single-photon detector, serving as trigger to herald the existence of the other photon. The heralded photon is injected into the Fredkin gate. The step-by-step description of preparation of input state $\ket{c}\ket{t_1}\ket{t_2}$ is 
\begin{figure*}[htp]
    \centering
    \includegraphics[width=0.4\textwidth]{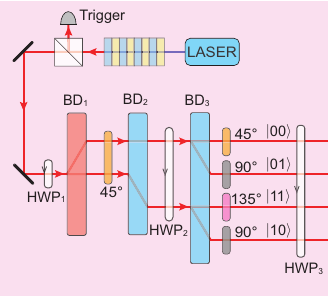}
    \caption{Schematic illustration of the generation of install states.}
    \label{fig:enter-label}
\end{figure*}
\begin{equation}\label{eq:state}
\begin{aligned}
\ket{H}&\xrightarrow{\text{HWP$_1$}@\theta_{1}} \cos2\theta_1\ket{H}+\sin2\theta_1\ket{V}
\xrightarrow{\text{BD1}} \cos2\theta_1\ket{Hl_0}+\sin2\theta_1\ket{Vl_1},\\
&\xrightarrow{\text{HWP}@45^\circ} 
\cos2\theta_1\ket{Vl_0}+\sin2\theta_1\ket{Hl_1}\xrightarrow{\text{BD2}} 
\cos2\theta_1\ket{Vl_0}+\sin2\theta_1\ket{Hl_2},\\
&\xrightarrow{\text{HWP$_2$}@\theta_2} \cos2\theta_1\sin2\theta_2\ket{Hl_0}-\cos2\theta_1\cos2\theta_2\ket{Vl_0}+\sin2\theta_1\cos2\theta_2\ket{Hl_2}+\sin2\theta_1\sin2\theta_2\ket{Vl_2},\\
&\xrightarrow{\text{BD3}}-\cos2\theta_1\cos2\theta_2\ket{Vl_0}+\cos2\theta_1\sin2\theta_2\ket{Hl_1}+\sin2\theta_1\sin2\theta_2\ket{Vl_2}+\sin2\theta_1\cos2\theta_2\ket{Hl_3},\\
&\xrightarrow{\text{HWPs on four modes}}\ket{H}\left(\cos2\theta_1\cos2\theta_2\ket{l_0}+\cos2\theta_1\sin2\theta_2\ket{l_1}+\sin2\theta_1\sin2\theta_3\ket{l_2}+\sin2\theta_1\cos2\theta_3\ket{l_3}\right),\\
&=\ket{H}\left(\cos2\theta_1\cos2\theta_2\ket{00}+\cos2\theta_1\sin2\theta_2\ket{01}+\sin2\theta_1\sin2\theta_3\ket{11}+\sin2\theta_1\cos2\theta_3\ket{10}\right)\\
&=\ket{H}\left(\cos2\theta_1\ket{0}+\sin2\theta_1\ket{1}\right)\left(\cos2\theta_2\ket{0}+\sin2\theta_2\ket{1}\right),\\
&\xrightarrow{\text{HWP}_3@\theta_3}\left(\sin2\theta_3\ket{H}+\cos2\theta_3\ket{V}\right)\left(\cos2\theta_2\ket{0}+\sin2\theta_2\ket{1}\right)\left(\cos2\theta_1\ket{0}+\sin2\theta_1\ket{1}\right).
\end{aligned}  
\end{equation}
In the characterizations of Fredkin gate, the input states are $\ket{000}$, $\ket{001}$, $\ket{010}$, $\ket{011}$, $\ket{100}$, $\ket{101}$, $\ket{110}$ and $\ket{111}$ for classical fidelity measurement, and $\ket{+10}$ for generation of three-qubit GHZ state. The angle setting of $[\theta_1, \theta_2, \theta_3]$ are listed in Table~\ref{tab:thetavalue1}.
\begin{table*}[h!]
\centering
    \begin{tabular}{c|c|c|c|c|c|c|c|c}
    \hline\hline
     $\ket{000}$& $\ket{001}$& $\ket{010}$&$\ket{011}$&$\ket{100}$&$\ket{101}$&$\ket{110}$&$\ket{111}$&$\ket{+10}$ \\
       \hline
       $\theta_{1}$ &0$^\circ$&0$^\circ$&0$^\circ$&0$^\circ$&45$^\circ$&45$^\circ$&45$^\circ$&45$^\circ$ \\
       \hline
       $\theta_{2}$&0$^\circ$&0$^\circ$&45$^\circ$&45$^\circ$&0$^\circ$&0$^\circ$&45$^\circ$&45$^\circ$ \\
       \hline
       $\theta_{3}$&0$^\circ$&45$^\circ$&0$^\circ$&45$^\circ$&0$^\circ$&45$^\circ$&0$^\circ$&45$^\circ$ \\
   \hline\hline
   \end{tabular}
   \caption{The setting of angles $\theta_1$, $\theta_2$ and $\theta_3$ to prepare input state.}
    \label{tab:thetavalue1}
\end{table*}

In the HS experiment, the state of control qubit is $\ket{+}$, and we set $\theta_3=22.5^\circ$. The angle settings of $[\theta_2, \theta_3]$ to prepare target states $\ket{t_1t_2}\in\{\ket{++}$, $\ket{+1}$, $\ket{+0}$, $\ket{1+}$, $\ket{11}$, $\ket{10}$, $\ket{0+}$, $\ket{01}$,  $\ket{00}\}$  are listed in Table~\ref{tab:thetavalue2}.
\begin{table*}[h!]
\centering
    \begin{tabular}{c|c|c|c|c|c|c|c|c|c}
    \hline\hline
     &$\ket{++}$& $\ket{+1}$& $\ket{+0}$&$\ket{1+}$&$\ket{11}$&$\ket{10}$&$\ket{0+}$&$\ket{01}$&$\ket{00}$ \\
       \hline
       $\theta_{1}$ &22.5$^\circ$&22.5$^\circ$&22.5$^\circ$&45$^\circ$&45$^\circ$&45$^\circ$&0$^\circ$&0$^\circ$&0$^\circ$ \\
       \hline
       $\theta_{2}$&22.5$^\circ$&45$^\circ$&0$^\circ$&22.5$^\circ$&45$^\circ$&0$^\circ$&22.5$^\circ$&45$^\circ$&0$^\circ$ \\

   \hline\hline
   \end{tabular}
   \caption{The setting of angles $\theta_1$ and $\theta_2$ to generate two-copy noisy state.}
    \label{tab:thetavalue2}
\end{table*}

\subsection{Measurement}
\begin{figure}
    \centering
\includegraphics[width=0.6\textwidth]{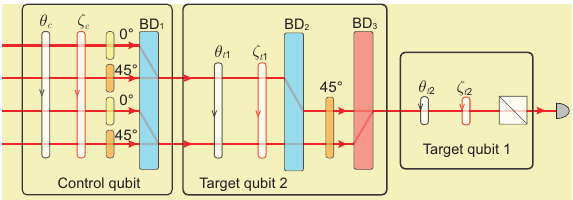}
    \caption{Illustration of setup to perfrom projective measurement on control qubit and two target qubits.}
    \label{fig:measure}
\end{figure}

The measurement setup is shown in Fig.~\ref{fig:measure}, which is able to perform the projective measurement on state $\ket{\varphi_c}\ket{\varphi_1}\ket{\varphi_2}=(\alpha_c\ket{0}+\beta_c\ket{1})\otimes(\alpha_1\ket{0}+\beta_1\ket{1})\otimes(\alpha_2\ket{0}+\beta_2\ket{1})$ with $\alpha, \beta, \alpha_1, \beta_1, \alpha_2, \beta_2$ being complex numbers. 
The step-by-step description of this projective measurement is

\begin{equation}\label{eq:measure}
\begin{aligned}
\ket{\varphi_c}\ket{\varphi_1}\ket{\varphi_2}&=(\alpha_c\ket{0}+\beta_c\ket{1})\otimes(\alpha_1\ket{0}+\beta_1\ket{1})\otimes(\alpha_2\ket{0}+\beta_2\ket{1})\\
&=(\alpha_c\ket{H}+\beta_c\ket{V})(\alpha_1\alpha_2\ket{00}+\alpha_1\beta_2\ket{01}+
\beta_1\alpha_2\ket{10}+
\beta_1\beta_2\ket{11})
\\
&=(\alpha_c\ket{H}+\beta_c\ket{V})(\alpha_1\alpha_2\ket{l_0}+
\alpha_1\beta_2\ket{l_1}+
\beta_1\beta_2\ket{l_2}+
\beta_1\alpha_2\ket{l_3})
\\
&\xrightarrow{\text{HWP}@\theta_c,\text{QWP}@\zeta_c}\ket{H} (\alpha_1\alpha_2\ket{l_0}+
\alpha_1\beta_2\ket{l_1}+
\beta_1\beta_2\ket{l_2}+
\beta_1\alpha_2\ket{l_3})\\
&\xrightarrow{\text{HWPs}@0^\circ,45^\circ,0^\circ,45^\circ}\ket{H} (\alpha_1\alpha_2\ket{l_0})+\ket{V} (\alpha_1\beta_2\ket{l_1})+\ket{H} (\beta_1\beta_2\ket{l_2})+\ket{V} (\beta_1\alpha_2\ket{l_3})\\
&\xrightarrow{\text{BD$_{1}$}}
(\alpha_2\ket{H}+\beta_2\ket{V})\alpha_1\ket{l_1}+(\alpha_2\ket{V}+\beta_2\ket{H})\beta_1\ket{l_3}\\
&\xrightarrow{\text{HWP}@\theta_{t2},\text{QWP}@\zeta_{t2}}\alpha_1\ket{H}\ket{l_1}+\beta_1\ket{V}\ket{l_3}\\
&\xrightarrow{\text{BD$_{2}$},\text{HWP}@45^\circ}\alpha_1\ket{V}\ket{l_2}+\beta_1\ket{H}\ket{l_3}\\
&\xrightarrow{\text{BD$_{3}$}}\alpha_1\ket{V}+\beta_1\ket{H}\\
&\xrightarrow{\text{HWP}@\theta_{t1},\text{QWP}@\zeta_{t1}}\ket{H}\
\end{aligned}  
\end{equation}

In our experiment, the projective measurement is performed on Pauli bases and the angle setting are listed in Table~\ref{tab:thetavalue3}.
\begin{table*}[h!]
\centering
    \begin{tabular}{c|c|c|c|c|c|c}
    \hline\hline
       $\ket{\varphi_c}$& $\ket{V}$& $\ket{H}$ &$\ket{+}$ & $\ket{-}$ & $\ket{R}$ &$\ket{L}$\\
       \hline
       $\theta_c$&0$^\circ$& 45$^\circ$&22.5$^\circ$&
       67.5$^\circ$&
      0$^\circ$&0$^\circ$\\
       \hline
              $\zeta_c$& 0$^\circ$& 0$^\circ$ & 0$^\circ$ & 0$^\circ$& 45$^\circ$&-45$^\circ$ \\
       \hline  \hline
       $\ket{\varphi_2}$& $\ket{0}$& $\ket{1}$ &$\ket{+}$ & $\ket{-}$ & $\ket{R}$ &$\ket{L}$\\
       \hline
       $\theta_{t2}$&0$^\circ$& 45$^\circ$&22.5$^\circ$&
       67.5$^\circ$&
      0$^\circ$&0$^\circ$\\
              \hline
       $\zeta_{t2}$&0$^\circ$& 0$^\circ$&0$^\circ$&
       0$^\circ$&
      45$^\circ$&-45$^\circ$\\
         \hline\hline      
       $\ket{\varphi_1}$& $\ket{0}$& $\ket{1}$ &$\ket{+}$ & $\ket{-}$ & $\ket{R}$ &$\ket{L}$\\
              \hline
      $\theta_{t1}$&45$^\circ$& 0$^\circ$&22.5$^\circ$&
       67.5$^\circ$&
      0$^\circ$&0$^\circ$\\
              \hline
        $\zeta_{t2}$&0$^\circ$& 0$^\circ$&0$^\circ$&
       0$^\circ$&
      45$^\circ$&-45$^\circ$\\
   \hline\hline
   \end{tabular}
   \caption{The setting of angles HWP and QWP to perform projective measurements of Pauli bases on control qubit, the first and second target qubits. For the first and second target qubits, $\ket{H(1)}$ and $\ket{V(0)}$ are the eigenbasis of Pauli-$Z$, $\ket{\pm}=\frac{1}{\sqrt{2}}(\ket{H(1)}\pm\ket{V(0)})$ are the eigenbasis of Pauli-$X$ and $\ket{R}=\frac{1}{\sqrt{2}}(\ket{H(1)}+i\ket{V(0)})$ and $\ket{L}=\frac{1}{\sqrt{2}}(\ket{H(1)}-i\ket{V(0)})$ are the eigenbasis of Pauli-$Y$. }
    \label{tab:thetavalue3}
\end{table*}
\subsection{Characterization of Fredkin gate}
\begin{figure*}
    \centering
    \includegraphics{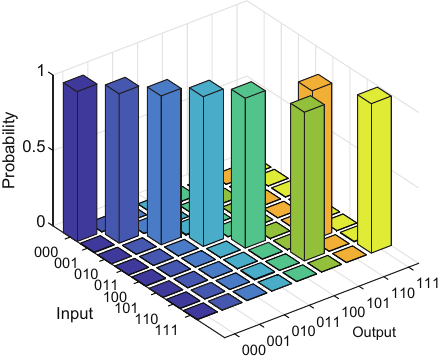}
    \caption{Measurement results of the Fredkin gate in computational basis.}
    \label{fig:truth}
\end{figure*}

We measure the truth table of Fredkin gate, which are the probabilities $P_{ijk|ct_1t_2}$ of the outcome state in the computational basis $\ket{ijk}$ with input state $\ket{ct_1t_2}\in\{\ket{000}, \ket{001}, \ket{010}, \ket{011}, \ket{100}, \ket{101}, \ket{110}, \ket{111}\}$, respectively. The measurement results are shown in Fig.~\ref{fig:truth} and the values of $P_{ijk|ct_1t_2}$ are listed in Table~\ref{tab:truth}. Then, we can calculate the classical fidelity $F_{zzz}$ of the Fredkin gate~\cite{Ono2017SR}
\begin{equation}
    F_{zzz}=\frac{1}{8}\left(P_{000|000}+P_{001|001}+P_{010|010}+P_{011|011}+P_{100|100}+P_{110|101}+P_{101|110}+P_{111|111}\right),
\end{equation}
and obtain $F_{zzz}=0.9898\pm0.0003$.
\begin{table*}[h!]
\centering
    \begin{tabular}{c|c|c|c|c}
    \hline\hline
   \diagbox{output}{input} & 000 & 001 & 010 & 011 \\
   \hline
   000&$0.976\pm0.009$& $0.0029\pm0.0005$&$0.0005\pm0.0002$&$0.00008\pm0.00008$\\
   \hline
   001&$0.0054\pm0.0006$&$0.990\pm0.009$&$0.0008\pm0.0002$&$0.0017\pm0.0004$\\
   \hline
   010& $0.0011\pm0.0003$&$0.00008\pm0.00008$&$0.990\pm0.009$& $0.0018\pm0.0004$\\
   \hline
   011&$0.0011\pm0.0003$&$0.0020\pm0.0004$&$0.0061\pm0.0007$&$0.994\pm0.009$\\
   \hline
   100&$0.015\pm0.001$&$0.0007\pm0.0003$&$0.00008\pm0.00008$&$0.0002\pm0.0002$\\
   \hline
   101&$0.0002\pm0.0002$&$0.0020\pm0.0004$&$0.00016\pm0.00008$&$0.0004\pm0.0002$\\
   \hline
   110&$0.00008\pm0.00008$&$0.0016\pm0.0004$&$0.0026\pm0.0005$&$0.0013\pm0.0003$\\
   \hline
   111&$0.0006\pm0.0002$&$0.0011\pm0.0003$&$0.00008\pm0.00008$&$0.0009\pm0.0002$\\
   \hline\hline
   \diagbox{output}{input} &100&101 &110&111 \\
   \hline000&$0.016\pm0.001$&$0.00008\pm0.00008$&$0.0002\pm0.0002$&$0.0016\pm0.0003$\\
   \hline
   001&$0.00008\pm0.00008$&$0.0009\pm0.0002$&$0.0021\pm0.0004$&$0.0010\pm0.0003$\\
   \hline
   010&$0.00008\pm0.00008$&$0.0014\pm0.0003$&$0.0004\pm0.0002$&$0.00008\pm0.00008$\\
   \hline
   011&$0.0024\pm0.0004$&$0.0008\pm0.0002$&$0.0003\pm0.0002$&$0.0029\pm0.0005$\\
   \hline
   100&$0.976\pm0.009$&$0.0002\pm0.0002$&$0.0019\pm0.0004$&$0.00008\pm0.00008$\\
   \hline
   101&$0.0039\pm0.0006$&$0.0003\pm0.0002$&$0.99\pm0.01$&$0.0007\pm0.0002$\\
   \hline
   110&$0.0005\pm0.0002$&$0.985\pm0.009$&$0.0010\pm0.0002$&$0.0010\pm0.0002$\\
   \hline
   111&$0.0010\pm0.0003$&$0.011\pm0.001$&$0.0007\pm0.0002$&$0.993\pm0.0009$\\
   \hline\hline
   \end{tabular}
   \caption{Truth tables of the Fredkin gate.}
    \label{tab:truth}
\end{table*} 

\begin{figure*}
    \centering
    \includegraphics{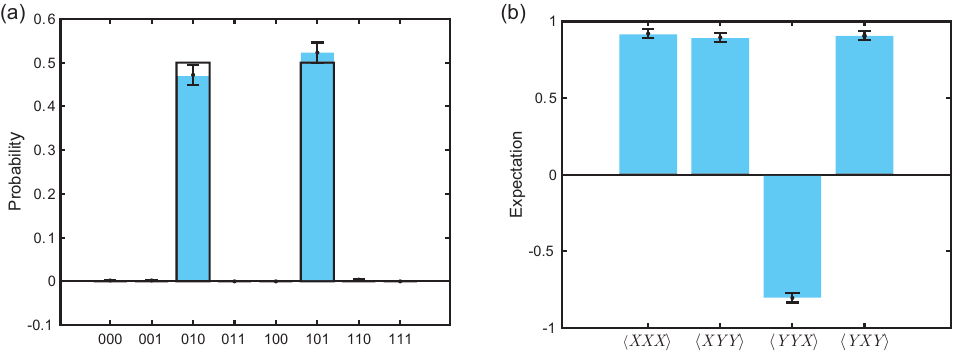}
    \caption{(a) Probabilities of the prepared GHZ state on computation bases, the black box represents the theoretical prediction of 0.5. (b) The expectation values of $\left<XXX\right>$, $\left<XYY\right>$, $\left<YYX\right>$, $\left<YXY\right>$.}
    \label{fig:proba}
\end{figure*}

We also generate a three-qubit GHZ state by inputting $\ket{+10}$ into the Fredkin gate. The output state is
\begin{equation}
\ket{\text{GHZ}_3}=\frac{1}{\sqrt{2}}(\ket{010}+\ket{101}).
\end{equation}
The fidelity of outcome state $\rho$ with respect to $\ket{\text{GHZ}_3}$ is $F=\tr\left(\ket{\text{GHZ}_3}\bra{\text{GHZ}_3}\rho\right)$, where $\ket{\text{GHZ}_3}\bra{\text{GHZ}_3}$ can be decomposed into
\begin{equation}
    \ket{\text{GHZ}_3}\bra{\text{GHZ}_3}=\frac{1}{2}\left(\ket{010}\bra{010}+\ket{101}\bra{101}\right)+\frac{1}{8}\left(XXX+XYY-YXY+YYX\right). 
\end{equation}
To calculate the fidelity $F$, we first measure the probabilities of outcome state on computational basis, and the results are shown in Fig.~\ref{fig:proba}(a). Then, we measure the expected values of observable $XXX$, $XYY$, $YXY$ and $YYX$, and the results are shown in Fig.~\ref{fig:proba}(b). Accordingly, we calculate the fidelity $F=0.933\pm0.002$. Using these data, we can estimate the process fidelity of Fredkin gate by~\cite{Ono2017SR}
\begin{equation}
    F_{\text{process}}=\frac{1}{2}F_{zzz}+\frac{1}{8}\left(\langle XXX\rangle+\langle XYY\rangle-\langle YXY\rangle+\langle YYX\rangle\right),
\end{equation}
and obtain $F_{\text{process}}=0.935\pm0.001$.

\section{More experimental results of distillation}
\begin{figure}[h]
    \centering
    \includegraphics[width=\linewidth]{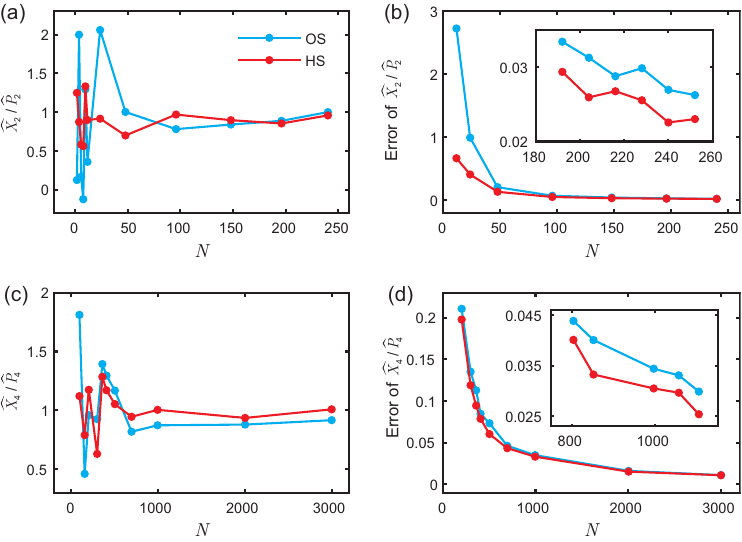}
    \caption{Experimental results of virtual distillation of expected value of observable $X$, with the estimator (a) $\widehat{X_2}/\widehat{P_2}$ and (c) $\widehat{X_4}/\widehat{P_4}$. Estimation error of (b) $\widehat{X_2}/\widehat{P_2}$ and (d) $\widehat{X_4}/\widehat{P_4}$ to corresponding target values. The red and blue dots represent the results with HS and OS, respectively.}
    \label{fig:distill34}
\end{figure}

\begin{figure}[h]
    \centering
    \includegraphics{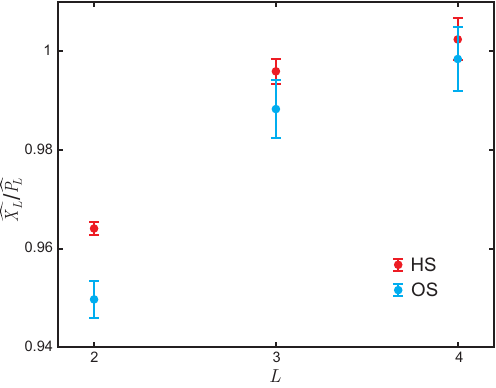}
    \caption{Experimental results of virtual distillation of expected value of observable $\langle X \rangle^{(L)}_\text{VD}$ with $L =2, 3$ and 4. The red and blue dots represent the results with HS and OS, respectively. }
    \label{fig:distillX234}
\end{figure}

The estimators of $\tr({O}\rho^2)$ and $\tr({O}\rho^4)$ with HS are
\begin{equation}
\widehat{O_2}_{(\text{HS})}=\frac{1}{M}\sum_{i\in \left[ M \right]}{\tr[O\widehat{\rho ^2}_{\left(i\right)}]}.
\end{equation}
\begin{equation}
\widehat{O_4}_{(\text{HS})}=\frac{1}{MM'}\sum_{i\in \left[ M \right] ,i^{\prime}\in \left[ M^{\prime} \right]}{\tr\left[ {O} \widehat{\rho ^2}_{\left( i \right)}\widehat{\rho^2 }_{\left( i^{\prime} \right)} \right] }.
\end{equation}
The results of $\langle X \rangle^{(2)}_\text{VD}$ and $\langle X \rangle^{(4)}_\text{VD}$ are shown in Fig.~\ref{fig:distill34}. Indeed, the larger $L$ is, the more accuracy $\langle X \rangle^{(L)}_\text{VD}$ is. To clarify this, we perform the VD of $\langle X \rangle^{(2)}_\text{VD}$, $\langle X \rangle^{(3)}_\text{VD}$ and $\langle X \rangle^{(4)}_\text{VD}$ with number of copies of $N=2400, 3600$ and $4800$ respectively, and the results are shown in Fig.~\ref{fig:distillX234}. The accuracy of $\langle X \rangle^{(L)}_\text{VD}$ increases with increasing $L$ either with HS and OS, however, the variance is smaller with HS compared to OS.   
\end{document}